\documentstyle[12pt]{article}
\def\abstract{\if@twocolumn
\section*{Abstract}
\else \small
\begin{center}
{ABSTRACT\vspace{-.5em}\vspace{0pt}}
\end{center}
\quotation
\fi}
\def\endabstract{\if@twocolumn\else\endquotation\fi}

\begin{document}

\centerline{\bf RENORMALIZATION GROUP AND SCALING WITHIN}
\centerline{\bf THE MICROCANONICAL FERMIONIC AVERAGE APPROACH}
\vspace{1.5 cm}

\centerline{V. AZCOITI, V. LALIENA}
\centerline{\em Departamento de F\'\i sica Te\'orica, Facultad
de Ciencias, Universidad de Zaragoza,}
\centerline{\em 50009 Zaragoza (Spain)}
\vspace{0.3cm}
\centerline{G. DI CARLO, A. F. GRILLO.}
\centerline{\em Istituto Nazionale di Fisica Nucleare,
Laboratori Nazionali di Frascati,}
\centerline{\em P.O.B. 13 - Frascati (Italy).}
\centerline{A. GALANTE}
\centerline{\em Dipartimento di Fisica dell'Universit\'a
dell'Aquila, 67100 L'Aquila, (Italy)}
\centerline{\em and Istituto Nazionale di Fisica Nucleare,
Laboratori Nazionali di Frascati,}
\centerline{\em P.O.B. 13 - Frascati (Italy).}

\begin{center}
\parbox{13.0cm}
{\begin{center} ABSTRACT \end{center}
{\small \hspace*{0.3cm}

The MFA approach for simulations with dynamical fermions in
lattice gauge theories allows in principle to explore
the parameters space of the theory (e.g. the $\beta, m$ plane
for the study
of chiral condensate in QED) without the need of computing the
fermionic determinant at each point.

We exploit this possibility for  extracting
both
the renormalization
group trajectories ("constant physics lines") and the
scaling function, and
we test it in the Schwinger Model.
We discuss the applicability of this method to realistic theories.
 }}
\end{center}

In this paper we present a further test of the potentialities of the
Microcanonical Fermionic Average (M.F.A.)\cite{GEN} method.
We apply this procedure to
the Schwinger Model.
Since in this approach, the main computer
cost resides in the evaluation of an effective fermionic action at
fixed
pure gauge energy by evaluating {\it all} the eigenvalues of the
fermionic
matrix at $m=0$, it is possible, at no extra
cost, to move in the plane $\beta, m$.

We have made full use of this fact for obtaining the
constant physics trajectories of the model, whithout any a priori
assumption of their form, which is exactly known in the Schwinger Model,
since here the Renormalization Group amounts to simple dimensional
analysis.

The continuum Schwinger model \cite{REUTER} is confining; it is
exactly solvable at zero fermionic mass, so
that we can compare the results of our simulations with
exact ones. It is superrenormalizable and
the Callan-Symanzik $\beta$ function vanishes.
The chiral current is
anomalous;
if the chiral limit is obtained from
$m \ne 0$, then the $\theta=0$ vacuum is selected.
In this vacuum the chiral condensate is (with one flavour)
$
 e_c^{-1}\langle \bar\psi\psi\rangle_c = 0.15995
$
This is the value of the chiral condensate to be compared with the
results
of lattice simulations, where its chiral limit is obtained from
$m \ne 0$.

The M.F.A. method is fully described in \cite{GEN}.
Starting from the partition function written
as
sum of the fermionic and
pure gauge contributions, we define the density of states at fixed
pure gauge Action ({\it i.e.} Euclidean Energy)
%\begin{equation}
%N(E)= \int DU \delta(S_G(U)-VE)
%\end{equation}
and an effective fermionic action
which is (minus) the logarithm of the average
of the fermionic determinant over fixed energy configurations.
All the thermodynamical quantities can be computed as 1-dimensional
integrals; bosonic operators can be directly computed at $m=0$
In the present simulation
 we use 1 flavour of
staggered fermions, and
the pure gauge part is described in terms of
{\it non compact} fields, so that the density of states is
analytically known
\cite{}.

 Since
$e_c$ is dimensionful, $\beta$ explicitely contains the lattice
spacing:
$\beta={1 \over a^2 e^2_c}$ so that the
continuum limit of the theory is approached, assuming
super-renormalizability, at $\beta \to \infty$.
The limit must be reached keeping fixed the
dimensionless ratio ${m_c \over e_c} = {\sqrt \beta} m$.
This ratio defines constant physics trajectories.
We present here the
results for the $100^2$ lattice.

We compute all the eigenvalues of the fermionic
matrix;
this allows us to compute exactly the fermionic determinant, and hence
the Effective Action for all values of
the mass,
including $m = 0$. We have done so for $8$ values of the energy,
from $0.02$ to $0.25$.

The chiral condensate
at $m=0$ vanishes on a finite lattice,
so it must be
obtained as the limit $m \to 0$. To reach the
correct continuum value,
this limit has to be taken simultaneously with the
$\beta \to \infty$ one,
keeping the product ${\sqrt \beta} m$ fixed.
This can be easily done with this method,
which does not require a separate
simulation of the fermionic contribution
for each pair of parameters $(\beta,m)$ \cite{prd}.

We have repeated this procedure for $10$ values of
$m_c / e_c$, always finding a clear  scaling window, where we
compute the chiral condensate.
Using a quadratic fit for the experimental points,
we obtain for the intercept:
$e_c^{-1} \langle \bar \psi \psi \rangle = 0.159 \pm 0.003 $
in perfect agreement with the theoretical value.

To better show  the potentialities of the method, we have investigated
the possibility of defining constant physics trajectories without
any a priori
knowledge of their behaviour. This procedure would be
the appropriate one in
any perturbatively renormalizable theory like QCD.

We identify as physical observables the continuum chiral
condensate and the
charge, and look for constant physics trajectories in the plane
$(\beta,m)$.

Since $ a \langle \bar \psi \psi \rangle_c =
  \langle \bar \psi \psi \rangle_L $, and that the plaquette
behaves as $1/\beta$ as $\beta \to \infty$: in this limit
$\langle E \rangle_L \to e^2_c a^2$. This implies that
\begin{equation}
R(\beta,m) =
{(\langle \bar \psi \psi \rangle_L)^2 \over \langle E \rangle_L}
\to
{(\langle \bar \psi \psi \rangle_c)^2\over e^2_c}
\end{equation}
is independent on the cutoff and then defines the constant physics
trajectories as the values in the
$\beta,m$ plane where R is constant.

Constant level lines on the surface defined by R
are thus the trajectories we
are looking for, and they are reported in Fig. 1, compared with
$\sqrt \beta m = constant$.

It can be seen that the agreement is indeed very good,
apart from large
masses and/or small $\beta$ where we are affected by
finite cutoff effects.

It is important to discuss also the applicability of
these results to more realistic
theories, for instance QCD.
The choice of the operators to be fixed in the
renormalization procedure is dictated by physical
motivations, and the
choice of the continuum charge is
intrinsic of the Schwinger model.

On the other hand, to
fully exploit the potentialities of the MFA method,
one is restricted to the
use of either bosonic operators, or fermionic ones
related to the
eigenvalues of the fermionic matrix.

In QCD a convenient choice would be the ratio of the
longitudinal to transverse
susceptibilities \cite{KOG}
$ R(\beta,m) = {\chi_L / \chi_T}$
where both numerator and denominator are computable from
the eigenvalues of
the fermionic matrix.

Once constant physics trajectories (which are
renormalization group
trajectories in renormalizable theories) have been defined,
any other physical
observable should be constant if computed along them.
For instance, if we compute hadronic masses in lattice units
$m_h a$, all their dependence  on bare parameters along
R.G. trajectories
will come from the lattice spacing $a$.
In this way one can extract the scaling functions.
This allows to extract the continuum values of
the observables.

This work has been partly supported
through a CICYT (Spain) - INFN (Italy)
collaboration.

\vskip 1.5 cm
\leftline{\bf FIGURE CAPTION}
\vskip 0.5 cm
\noindent
Constant Physics trajectories in ($\beta,m$) plane (continuous
lines); circles are $\beta^{1/2} m=constant$ lines.

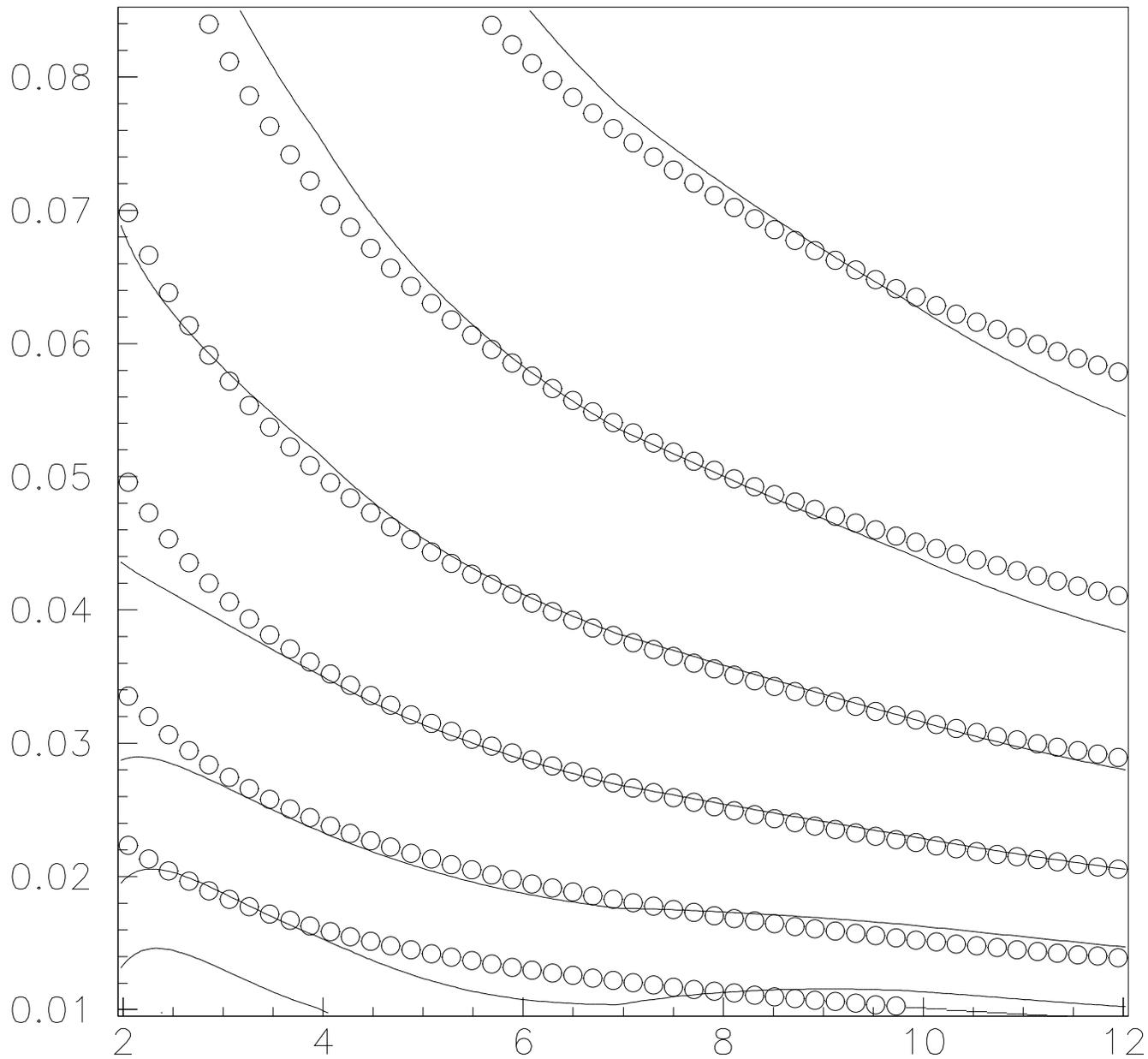
\end{document}